\documentclass[a4paper]{jpconf}
\usepackage{graphicx}
\usepackage{epstopdf}
\begin{document}
\title{On the validity of cosmic no-hair conjecture in an anisotropic inflationary model}

\author{Tuan Q Do}

\address{Faculty of Physics, VNU University of Science, Vietnam National University, Hanoi 120000, Vietnam}

\ead{tuanqdo@vnu.edu.vn}
\begin{abstract}
We will present main results of our recent investigations on the validity of cosmic no-hair conjecture proposed by Hawking and his colleagues long time ago in the framework of an anisotropic inflationary model proposed by Kanno, Soda, and Watanabe. As a result, we will show that the cosmic no-hair conjecture seems to be generally violated in the Kanno-Soda-Watanabe model for both canonical and non-canonical scalar fields due to the existence of a non-trivial coupling term between scalar and electromagnetic fields. However, we will also show that the validity of the cosmic no-hair conjecture will be ensured once a unusual scalar field called the phantom field, whose kinetic energy term is negative definite, is introduced into the Kanno-Soda-Watanabe model. 
\end{abstract}

\section{Introduction}\label{sec1}
An inflationary universe has been regarded as one of  leading paradigms of modern cosmology due to the fact that it has helped us to solve several important problems in cosmology such as the {\it monopole}, {\it horizon}, and {\it flatness} problems \cite{guth}. More interestingly, many theoretical predictions based on the  inflation paradigm have been well confirmed by some high technology satellites such as the Wilkinson Microwave Anisotropy Probe (WMAP) \cite{Komatsu:2010fb} and Planck \cite{Planck} built to observe the cosmic microwave background radiation (CMB). However, some anomalies detected in CMB such as  the hemispherical asymmetry and the Cold Spot \cite{Komatsu:2010fb,Planck} cannot be explained in the standard inflationary framework, which is based on the assumption that the spacetime of the early universe should be homogeneous and isotropic as the Friedmann-Lemaitre-Robertson-Walker (FLRW) metric. Therefore, some modifications to the metric or fields of standard inflationary models might be necessary. For example, the Bianchi metrics, which are homogeneous but anisotropic spacetimes \cite{bianchi}, might be useful  to realize the nature of the mentioned exotic features of CMB \cite{Ade:2013vbw}. It is worth noting that some theoretical predictions of  anisotropic inflationary models can be found in some earlier papers \cite{Pitrou:2008gk}. 

Note again that  the early universe might be slightly anisotropic according to the recent observations of WMAP and Planck. Hence, it is natural to think about the state of the late-time universe. Will it be isotropic or still be slightly anisotropic is an open question to all of us.  Theoretically, an important hint to this question can be obtained from the so-called cosmic no-hair conjecture proposed by Hawking and his colleagues long time ago  \cite{GH,Barrow:1987ia}. In particular, the statement of this conjecture is that our late-time universe would be simply homogeneous and isotropic, no matter the initial states and conditions of the early universe. However, a complete proof to this conjecture is not easy to achieve. Note that some partial proofs to this conjecture have been done in \cite{wald}. On the other hand, a counter-example to this conjecture has been claimed to exist  in a supergravity motivated model by Kanno, Soda, and Watanabe (KSW) recently \cite{MW}. For recent interesting reviews on this model, see \cite{SD}. Hence, it is important to examine the validity of the cosmic no-hair conjecture in the KSW model. In particular, we should check if the conjecture is still violated when a canonical scalar field is replaced by non-canonical scalar ones such as the Dirac-Born-Infeld (DBI) \cite{Silverstein:2003hf}, supersymmetric DBI (SDBI) \cite{Sasaki:2012ka}, and covariant Galileon fields \cite{G-inflation}. In addition, we should also check the fate of the conjecture when unusual field(s), e.g., the phantom field,  one of alternative solutions to the dark energy problem in cosmology  due to it negative kinetic term \cite{phantom}, is introduced into the KSW model. As a result, these scenarios have been investigated explicitly in our recent published papers \cite{WFK,WFK1,WFK2,WFK3,WFK4}. 

The present article will be devoted to summarize basic results of our published papers. It will be organized as follows. A short introduction of our study has been written in the section \ref{sec1}. Non-canonical extensions of KSW model will be briefly shown in the section \ref{sec2}. The role of phantom field to the validity of cosmic no-hair conjecture will be discussed in the section \ref{sec3}. Finally, conclusions  will be given in the section \ref{sec4}. 
\section{Non-canonical extensions of KSW model} \label{sec2}
\subsection{Basic setup of KSW model} 
Before going to present details of the non-canonical extensions of KSW model mentioned above, we would like to draw here a very brief setup of KSW model \cite{MW,SD} since it would be very useful in order to compare the KSW model with its non-canonical extensions. As a result, an action of KSW model  is given by \cite{MW,SD}
\begin{equation} \label{c2.1}
  S_{\rm KSW} = \int {d^4 } x\sqrt { - g}  \left[ {\frac{{M_p^2 }}{2}R - \frac{{1 }}
{2}\partial _\mu  \phi \partial ^\mu  \phi  - V\left( {\phi } \right) -\frac{1}{4} {f^2 \left( {\phi } \right) }{F_{\mu \nu } F^{\mu \nu }} } \right], 
\end{equation}
with $M_p$ is the reduced Planck mass, $\phi$ is a canonical scalar field, and ${F_{\mu\nu}\equiv \partial_\mu A_\nu -\partial_\nu A_\mu}$ is the field strength of the electromagnetic (vector) field $A_\mu$. In addition, $f(\phi)$ is a gauge kinetic function of $\phi \equiv \phi(t)$. Note that $f(\phi) =1$ is always chosen  in usual scenarios. As a result, varying this action with respect to $g^{\mu\nu}$ leads to the corresponding Einstein field equations:
\begin{equation}\label{c2.2}
 M_p^2\left( {R_{\mu \nu }  - \frac{1}
{2}Rg_{\mu \nu } } \right) - \partial _\mu  \phi \partial _\nu  \phi + g_{\mu \nu } \left[ \frac{1}
{2} \partial ^\sigma  \phi \partial _\sigma  \phi +   {V\left( {\phi } \right) + \frac{1}
{4}f^2 \left( {\phi  } \right)F^{\rho \sigma } F_{\rho \sigma } } \right] - f^2 \left( {\phi } \right)F_{\mu \gamma } F_\nu ^\gamma  = 0.
\end{equation}
Additionally, the field equations of vector and scalar fields read
\begin{eqnarray}\label{c2.3}
 \frac{\partial }
{{\partial x^\mu  }}\left[ {\sqrt { - g} f^2 \left( \phi \right)F^{\mu \nu } } \right] &=& 0,
\\  \label{c2.4}
 \ddot \phi     + 3H\dot \phi  + \frac{{\partial V\left( {\phi } \right)}}
{{\partial \phi }} + \frac{1}
{2}f\left( {\phi  } \right)\frac{{\partial f\left( {\phi } \right)}}
{{\partial \phi }}F_{\mu \nu } F^{\mu \nu } &=&0 ,
\end{eqnarray}
respectively. By taking the Bianchi type I metric,
\begin{equation} \label{c2.BImetric}
ds^2= - dt^2+ \exp\left[{2\alpha \left( t \right) - 4\sigma \left( t \right)} \right] dx^2 +\exp\left[{2\alpha \left( t \right) + 2\sigma \left( t \right)}\right] \left({dy^2+dz^2}\right),
\end{equation}
along with the vector and scalar fields chosen as $A_\mu   = \left( {0,A_x \left( t \right),0,0} \right)$ and $\phi  = \phi \left( t \right)$, we are able to define the corresponding solution for the equation of vector field (\ref{c2.3}) as
\begin{equation}\label{c2.5}
\dot A_x \left( t \right) = f^{ - 2} \left( \phi  \right) \exp\left[{ - \alpha  - 4\sigma }\right] p_A ,
\end{equation}
with $p_A$ is  a constant of integration. As a result, plugging this solution into the other field equations leads to
\begin{eqnarray}\label{c2.eq7}
\ddot \phi  &=&  - 3\dot \alpha \dot \phi  - \frac{{\partial V\left( {\phi } \right)}}
{{\partial \phi }} + f^{ - 3} \left( \phi  \right)\frac{{\partial f\left( {\phi  } \right)}}
{{\partial \phi }}\exp\left[{ - 4\alpha  - 4\sigma } \right]p_A^2 ,\\
\label{c2.eq9}
 \dot \alpha ^2  &=& \dot \sigma ^2  + \frac{{1 }}
{3M_p^2}\left[ {\frac{1}
{2}\dot \phi ^2 + V\left( \phi \right) + \frac{1}
{2}f^{ - 2} \left( \phi\right)\exp\left[{ - 4\alpha  - 4\sigma } \right]p_A^2 } \right] , 
\\ \label{c2.eq10}
\ddot \alpha  & = & - 3\dot \alpha ^2  + \frac{{1 }}
{M_p^2} V\left( \phi \right) + \frac{{1}}
{6M_p^2}f^{ - 2} \left( \phi  \right)\exp\left[{ - 4\alpha  - 4\sigma } \right]p_A^2 , 
\\ \label{c2.eq11}
 \ddot \sigma & =&  - 3\dot \alpha \dot \sigma  + \frac{{1 }}
{3M_p^2}f^{ - 2} \left( \phi\right)\exp\left[{ - 4\alpha  - 4\sigma } \right]p_A^2 ,
\end{eqnarray}
 Now, by choosing  the exponential potential and gauge kinetic function,
\begin{equation}
V(\phi) = V_{0} \exp\left[{\frac{\lambda}
{{M_p }}\phi }\right]; ~f\left( \phi  \right) = f_0 \exp \left[{\frac{\rho }{{M_p }}\phi  }\right],
\end{equation}
along with the following ansatz,
 \begin{equation}
\alpha = {\zeta} \log \left( t \right); ~\sigma  = {\eta} \log \left( t \right); ~\frac{\phi }
{{M_p }}  = \xi \log \left( t \right) + \phi _0,
\end{equation}
ones have been able to define the corresponding solutions for scale factors of metric from the above field equations such as \cite{MW}
\begin{equation}
{ \zeta} = \frac{\lambda^2 + 8 \rho \lambda + 12 \rho^2 +8}{6\lambda (\lambda + 2\rho)};~{\eta} = \frac{\lambda^2 + 2\rho \lambda -4 }{3\lambda (\lambda + 2\rho)}.
\end{equation}
Here $\lambda$, $\rho$, $V_0$, $f_0$, and $\phi_0$ are positive constants. It is noted that $\sigma$ acts as a deviation from the isotropy, i.e., $\sigma \ll \alpha$ in order to be consistent with the data of WMAP and Planck. Consequently, $\eta \ll \zeta$ is required for the obtained anisotropic inflationary solution. As a result, this constraint will be easily fulfilled if $\rho \gg \lambda \sim {\cal O}(1)$. By converting the field equations into autonomous equations of  dynamical variables defined as 
\begin{equation}
{X}= \frac{\dot\sigma}{\dot\alpha}; ~{Y}=\frac{\dot\phi}{M_p \dot\alpha}; ~ {Z} =\frac{1}{f_0 M_p \dot\alpha} \exp \left [-\frac{\rho }{M_p }\phi-2\alpha-2\sigma \right] p_A,
\end{equation}
ones have shown that the anisotropic power-law solution found above is indeed equivalent to an anisotropic fixed point of dynamical system, which is a non-trivial solution of $dX/d\alpha =dY/d\alpha=dZ/d\alpha=0$,
\begin{eqnarray}
{X} &=& \frac{2 \left( \lambda^2 +2\rho \lambda -4 \right)}
            {\lambda^2 + 8 \rho \lambda + 12 \rho^2 +8};~ {Y}= - \frac{12 \left( \lambda +2\rho  \right)}
            {\lambda^2 + 8 \rho \lambda + 12 \rho^2 +8}, \nonumber\\
{Z^2} &=&  \frac{ 18 \left( \lambda^2 +2\rho \lambda -4 \right)
             \left(-\lambda^2 + 4\rho \lambda +12 \rho^2 +8\right) }
            {\left( \lambda^2 + 8 \rho \lambda + 12 \rho^2 +8 \right)^2}  .  
\end{eqnarray}
Moreover, ones have shown that this anisotropic fixed point, or the corresponding anisotropic power-law solution, is indeed a stable and attractor solution during the inflationary phase \cite{MW}. This implies that  the late-time universe in the KSW model will be anisotropic rather than isotropic, meaning that  the cosmic no-hair conjecture is violated in the KSW model. Interested readers are encouraged to read papers in \cite{MW} for more technical details. Additionally, cosmological implications of KSW model, e.g., imprints of anisotropic inflation on the CMB through correlations between $T$, $E$, and $B$ modes, can be found in the recent interesting review papers in \cite{SD}. In the following subsections, we will present the non-canonical extensions of KSW model mentioned in the introduction section to see whether the cosmic no-hair conjecture holds. 
\subsection{DBI model}
In this subsection, we will present basic results of an interesting extension of KSW model, the DBI model \cite{Silverstein:2003hf}, in which the canonical scalar field $\phi$ is replaced by the  non-canonical DBI one \cite{WFK1} as follows 
\begin{equation}\label{c3.DBI}
 S_{\rm DBI} =  \int {d^4 } x\sqrt {- g} \left[ {\frac{{M_p^2}}
{2}R +{ \frac{1}
{{f\left( \phi  \right)}}\frac{\gamma-1}{\gamma}}- V\left( \phi  \right) - \frac{1}
{4}h^2 \left( \phi  \right)F_{\mu \nu } F^{\mu \nu } } \right] ,
\end{equation}
with $\gamma \equiv 1/\sqrt {1 +f\left( \phi  \right)  \partial _\mu  \phi \partial ^\mu  \phi}$  is the Lorentz factor characterizing the motion of the D brane \cite{Silverstein:2003hf}. It is clear that $S_{\rm DBI}$ will reduce to $S_{\rm KSW}$ once $\gamma \to 1$ (or equivalently $f(\phi)\to 0$). It appears that $\gamma \ge 1$ for non-negative $f(\phi)$ if $\phi=\phi(t)$. Note that $h(\phi)$ in this model is identical to $f(\phi)$ in the KSW model. As a result, the following field equations turn out to be \cite{WFK1}
\begin{eqnarray} \label{c3.eqnphi}
\ddot \phi  &=&  - \frac{{3\dot \alpha }}
{{\gamma ^2 }}\dot \phi  - \frac{{\partial _\phi V }}
{{\gamma ^3 }} - \frac{{\partial _\phi f }}
{{2f}}\frac{{\left( {\gamma  + 2} \right)\left( {\gamma  - 1} \right)}}
{{\left( {\gamma  + 1} \right)\gamma }}\dot \phi ^2  + \frac{h^{ - 3} \partial _\phi h}{\gamma^3} \exp \left[{ - 4\alpha  - 4\sigma }\right] p_A^2 ,\\
\label{c3.eqnhamiltion}
 \dot \alpha ^2 & =& \dot \sigma ^2  + \frac{{1 }}
{3}\left[ {\frac{\gamma^2}
{\gamma+1}\dot \phi ^2  + V\left( \phi  \right) + \frac{1}
{2}h^{ - 2} \exp \left[{ - 4\alpha  - 4\sigma }\right] p_A^2 } \right] ,
\\ \label{c3.eqnalpha}
\ddot \alpha & = & - 3\dot \alpha ^2  +  \left[ {\frac{\gamma \left({\gamma-1}\right)}
{2\left({\gamma+1}\right)}\dot \phi ^2 + V\left( \phi  \right) }\right] +\frac{{1}}
{6}h^{ - 2}\exp \left[{ - 4\alpha  - 4\sigma }\right] p_A^2 , 
\\  \label{c3.eqnsigma}
\ddot \sigma & = & - 3\dot \alpha \dot \sigma  + \frac{{1 }}
{3}h^{ - 2}\exp \left[{ - 4\alpha  - 4\sigma }\right] p_A^2 ,
\end{eqnarray}
here we have used the same setup for fields and metric as proposed in the KSW model. As a result, the corresponding power-law anisotropic solution turns out to be  \cite{WFK1}
\begin{equation}
{\zeta}  = \frac{\lambda ^2  + 8\rho \lambda  + 12\rho ^2  + 8\gamma _0}{6\lambda \left( {\lambda  + 2\rho } \right)};~ {\eta}   = \frac{{\lambda ^2  + 2\rho \lambda  - 4\gamma_0}}
{{3\lambda \left( {\lambda  + 2\rho } \right)}}. \nonumber
\end{equation}
which is equivalent with an anisotropic fixed point of dynamical system defined as
\begin{eqnarray}
{X} &=&\frac{2\left[{\hat\gamma_0 \lambda\left({\lambda+2\rho}\right) -4}\right]}{\hat\gamma_0\left({\lambda^2+8\lambda\rho+12\rho^2}\right)+8};~ {Y} =-\frac{12\hat\gamma_0\left({\lambda+2\rho}\right)}{\hat\gamma_0\left({\lambda^2+8\lambda\rho+12\rho^2}\right)+8}, \nonumber\\ 
{Z^2} &=&\frac{18\left[{\hat\gamma_0 \lambda\left({\lambda+2\rho}\right)-4}\right]\left[{\hat\gamma_0\left({-\lambda^2+4\lambda\rho+12\rho^2}\right)+8}\right]}{\left[{\hat\gamma_0\left({\lambda^2+8\lambda\rho+12\rho^2}\right)+8}\right]^2}; ~{\hat\gamma_0 =\gamma_0^{-1}},
\end{eqnarray}
here $\gamma$ has been fixed to be a constant, i.e., $\gamma=\gamma_0 \geq 1$. It is clear that these solutions will recover that of KSW model shown above in the limit $\gamma_0 \to 1$. Of course, $\rho \gg \lambda \sim {\cal O}(1)$ is also required to have the anisotropic inflationary solution, similar to the KSW model. And using the dynamical system approach, we have been able to show that the anisotropic power-law solution of DBI model is also stable and attractive as that of KSW model during the inflationary phase \cite{WFK1} (see the plot listed below for more details). Hence, we can conclude that the cosmic no-hair conjecture is also violated in the DBI extension of KSW model.
\subsection{SDBI model}
As a result, an action of SDBI extension of KSW model is given by \cite{Sasaki:2012ka,WFK2}
\begin{equation} 
S_{\rm SDBI} = \int {d^4 } x\sqrt {-g} \left[ {\frac{{M_p^2}}
{2} R+ \frac{1}
{{f\left( \phi  \right)}}\frac{\gamma-1}{\gamma} - {\Sigma_0^2}~U\left( \phi  \right) - \frac{1}
{4}h^2 \left( \phi  \right)F_{\mu \nu } F^{\mu \nu } } \right],
\end{equation}
with 
\begin{equation}
{\Sigma_0}(\gamma) =\left({\frac{\gamma+1}{2\gamma}}\right)^{1/3} \le 1;~ U(\phi) = \left(\frac{27 }{2f(\phi)}\right)^{\frac{1}{3}} \left(\frac{d {W}(\phi)}{d\phi}\right)^{\frac{4}{3}}.
\end{equation}
Here $W(\phi)$ plays as  a {super-potential}. For a detailed derivation of this action, one can see \cite{Sasaki:2012ka}. As a result, the corresponding field equations of SDBI model are given by \cite{WFK2}
\begin{eqnarray}
\Sigma_1 \ddot \phi  &= &  - 3 \Sigma_2 \dot \alpha \dot \phi  -\left({\Sigma_0^{2}+\frac{\gamma}{3\Sigma_0}f \dot\phi^2}\right)\partial_\phi U  -\frac{\Sigma_1-\gamma^3}{2}\frac{{\partial_\phi f }}{{f }}\dot\phi^2  
- \frac{{\partial_\phi f }}
{{2f }}\frac{{\left( {\gamma  + 2} \right)\left( {\gamma  - 1} \right)\gamma^2}}{{\gamma  + 1 }}\dot \phi ^2 \nonumber \\
&& + h^{ - 3}   \partial _\phi h  \exp \left[{ - 4\alpha  - 4\sigma }\right] p_A^2 , \\
\dot \alpha ^2  &=& ~ \dot \sigma ^2  + \frac{{1 }}{3} \left[  \frac{\gamma^2}
{\gamma+1}\dot \phi ^2   +\frac{2\gamma+1}{3} \Sigma_0^2 U  + \frac{h^{ - 2}}
{2} \exp \left[{ - 4\alpha  - 4\sigma }\right] p_A^2  \right] , \\
\ddot \alpha & = & ~ - 3\dot \alpha ^2  +\frac{\gamma \left({\gamma-1}\right)}
{2\left({\gamma+1}\right)}\dot \phi ^2  + \frac{\gamma+2}{3} \Sigma_0^2 U  +\frac{{h^{ - 2}}}
{6}\exp \left[{ - 4\alpha  - 4\sigma }\right] p_A^2 , \\
\ddot \sigma  &= & ~ - 3\dot \alpha \dot \sigma  + \frac{{h^{ - 2} }}
{3}\exp \left[{ - 4\alpha  - 4\sigma }\right] p_A^2 .
\end{eqnarray}
with
\begin{equation}  \label{sig1}
\Sigma_1 = \gamma^3 + \frac{\gamma}{9\Sigma_0} \left({3\gamma^2+\gamma-1}\right)f U;~
\Sigma_2 = \gamma + \frac{\gamma}{3\Sigma_0} f U .
\end{equation}
Hence, the following Bianchi type I power-law solution of this model turns out to be \cite{WFK2}
\begin{eqnarray}
{\zeta} &=& \frac{N- \sqrt{N^2-4MP}}{2M}; ~{\eta}=-{\zeta} +\frac{\rho}{\lambda}+\frac{1}{2}, \\
M&=&18\lambda^2\left({\gamma_0^2-1}\right);~N=3\lambda \left({\gamma_0+1}\right) \left[{ \left({5\gamma_0+1}\right)\lambda+6 \left({\gamma_0+1}\right)\rho}\right] , \\
P&=& \left(\gamma_0+1\right)\left[ \left(2\gamma_0 +1\right)  \lambda^2 +2  \left(5\gamma_0 +7 \right) \lambda \rho+12\left(\gamma_0  +2 \right) \rho^2  \right]+8\gamma_0\left(5\gamma_0 +1 \right) . 
\end{eqnarray}
Similar to the DBI model, we are able to define the following anisotropic fixed point of SDBI model to be
\begin{equation}
X= \frac{-S_2-\sqrt{S_2^2+16S_1S_3}}{4S_1};~Y=-\frac{4}{\lambda+2\rho}\left(X+1\right);~
Z^2=3X\left(1-2X\right)-3\rho X Y,
\end{equation}
with
\begin{eqnarray}
S_1 &=& \left(\hat\gamma_0 +1 \right) \left[ \left(\hat\gamma _0+2\right)\lambda^2+2\left(7\hat\gamma_0+5\right)\lambda\rho+12\left(2\hat\gamma_0+1\right)\rho^2 \right]+8\left(\hat\gamma_0+5\right),\\
S_2 &=& \left(\hat\gamma_0+1\right) \left[\left(\hat\gamma_0-7\right)\lambda^2+8\left(4\hat\gamma_0-1\right)\lambda\rho +12 \left(5\hat\gamma_0+1\right)\rho^2 \right] +32 \left(\hat\gamma_0+5\right),\\
S_3 &=& \left(\hat\gamma_0+1\right)\left[ \left(5\hat\gamma_0+1\right)\lambda^2 +4\left(4\hat\gamma_0-1 \right)\lambda\rho +12\left(\hat\gamma_0-1\right)\rho^2\right]- 8\left(\hat\gamma_0+5\right).
\end{eqnarray}
Here $\gamma =\gamma_0 =$  constant and $\hat\gamma_0 =\gamma_0^{-1}$. It is noted that $\gamma_0$ in the SDBI model has been required to obey the following constraint \cite{WFK2} 
\begin{equation}
1\leq \gamma_0 < 1+\frac{\lambda}{\rho}.
\end{equation}
in order to have the anisotropic power-law solution. This result is consistent with the investigation for isotropic solutions done in \cite{Sasaki:2012ka}. Note again that $\gamma_0$ can be arbitrarily larger than one in the DBI extension of KSW model \cite{WFK1}. In addition, $\gamma$ in the DBI model is expected to be  larger than one in order to have the significant primordial non-Gaussianity in CMB \cite{Silverstein:2003hf}. However, the observations of WMAP and Planck have not confirmed the existence of the CMB  non-Gaussianity. Hence, $\gamma \simeq 1$ is cosmologically reasonable.
  For the stability of the obtained anisotropic power-law inflationary solution, we have used the dynamical system approach to investigate and arrived at a conclusion that it is stable and attractor solution, similar to that found in the KSW and DBI models \cite{WFK2}. For instance, one can see the plot listed below for numerical results showing  the attractor behavior of the obtained solution.
\subsection{Galileon model}
Next, we will consider another interesting extension of KSW model \cite{WFK4}, in which both kinetic and potential terms of the canonical scalar field are replaced by non-canonical terms of the so-called covariant Galileon model \cite{G-inflation} as follows
\begin{equation} 
S _{\rm G}= \int {d^4 } x\sqrt { -g}\left\{ \frac{{M_p^2 }}
{2}R + {k_0 \exp \left[ {\frac{{\tau \phi }}
{{M_p }}} \right]X} - {g_{0} \exp \left[ {\frac{{\lambda \phi }}
{{M_p }}} \right]X \opensquare \phi}  
- \frac{f_0^2}
{4}\exp \left[ {-\frac{{2\rho \phi }}{{M_p }}} \right]F_{\mu \nu } F^{\mu \nu } \right\}, 
\end{equation}
here $X \equiv -\frac{1}{2}\partial^\mu \phi \partial_\mu \phi $.
As a result, the corresponding field equations of this model turn out to be \cite{G-inflation,WFK4}
\begin{eqnarray}
&&  {\cal E}^{(2)}+{\cal E}^{(3)}+ f^{ - 3} f_{\phi } \exp \left[ {-4\alpha  -4\sigma } \right] p_A^2 =0 ,\\
&&-3M_p^2 \left( {\dot \alpha ^2  - \dot \sigma ^2 } \right) + \frac{k_0}{2} \exp\left[{\frac{\tau\phi}{M_p}}\right] \dot\phi^2 + 3g_0\exp\left[{\frac{\lambda\phi}{M_p}}\right] \dot\alpha \dot \phi ^3  - \frac{g_0\lambda}{2M_p}\exp\left[{\frac{\lambda\phi}{M_p}}\right]  \dot\phi^4 \nonumber\\
&& +\frac{f^{ - 2}}{2}  \exp \left[ {-4\alpha  -4\sigma } \right] p_A^2 =0, \\
 &&-M_p^2\ddot\alpha-3M_p^2 \dot\alpha^2 + \frac{g_0}{2}\exp\left[{\frac{\lambda\phi}{M_p}}\right]\left({\ddot\phi+3\dot\alpha \dot\phi}\right)\dot\phi^2 +\frac{f^{ - 2}}{6 }  \exp \left[ {-4\alpha  -4\sigma } \right] p_A^2=0,\\ 
&&\ddot\sigma+3\dot\alpha \dot\sigma - \frac{f^{ - 2}}{3 M_p^2}  \exp \left[ {-4\alpha  -4\sigma } \right] p_A^2 =0.
\end{eqnarray}
where 
\begin{eqnarray} 
 \label{eq.E2.simp}
{\cal E}^{( 2 )} &=& k_0 \left({\opensquare \phi  - \frac{\tau}{M_p}X}\right)  \exp \left[ {\frac{{\tau \phi }}
{{M_p }}} \right] ,\\ 
{\cal E}^{( 3 )} & =& g_{0}\left\{{ \frac{\lambda^2}{M_p^2}\frac{\dot\phi^2}{2}  -  \left[ {6H\frac{\ddot\phi}{\dot\phi} +9H^2 - \left({\sum\limits_{i = 1}^3 {{H_{i }^2} } }\right)- R_{0 0 }} \right]    + \frac{2\lambda}{M_p} \ddot\phi  }\right\} \dot\phi^2 \exp \left[ {\frac{{\lambda \phi }}
{{M_p }}} \right].
\end{eqnarray}
It is noted that $H \equiv (H_1 +H_2 +H_3) /3$ is the mean Hubble parameter along with its components $H_i \equiv \dot a_i/a_i $ ($i=1-3$). As a result, the corresponding solutions of this model can be defined to be \cite{WFK4}
\begin{equation}
\zeta = \frac{5}{12}+\frac{z}{2} + \frac{\sqrt{\Delta}}{12};~  \eta=-\zeta+ z +  \frac{1}{2},
\end{equation}
with 
\begin{equation}
z\equiv \frac{\rho}{\lambda}; ~\Delta\equiv 9-\frac{64 k_0}{\lambda^2}  -60z^2-20z,
\end{equation}
provided that $k_0 \le - 3 \lambda ^2 z^2/4$. Similar to the KSW model and its (S)DBI extensions, $\rho$ must be much larger than $\lambda$ to ensure the existence of inflationary solutions \cite{WFK4}. As a result, if $\rho \gg \lambda$ we will have the following approximations such as
\begin{equation}
\zeta \simeq z \gg 1;~ \eta \simeq \frac{1}{12};~ k_0 \simeq -\frac{3\rho^2}{2}.
\end{equation}
Additionally, the corresponding anisotropic fixed point of this model turns out to be
\begin{equation}
X = \frac{ -12z^2 +8 z +7 -32 \frac{k_0}{\lambda^2} - 3 \left(2z+1\right) \sqrt{\Delta}}{8 \left(6z ^2+5 z +1+4 \frac{k_0}{\lambda^2}\right)};~Y=\frac{4\left(X+1\right)}{\lambda+2\rho};~Z^2 =- 3X \left(2X -\rho  Y -1  \right) .
\end{equation}
Furthermore, the stability analysis and numerical result have confirmed that the anisotropic power-law inflationary solution of Galileon model is indeed stable and attractive as expected. Indeed, one can see the Fig. 1 for the numerical confirmation.
\begin{figure}
\begin{center}
{\includegraphics[height=50mm]{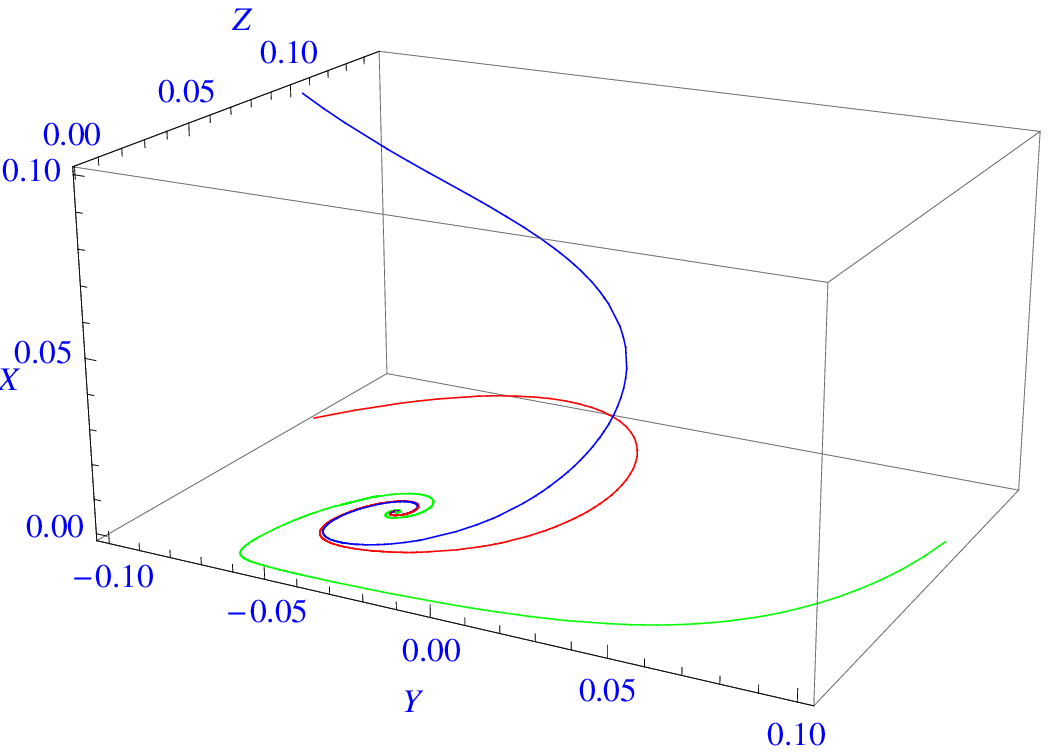}}\quad
{\includegraphics[height=40mm]{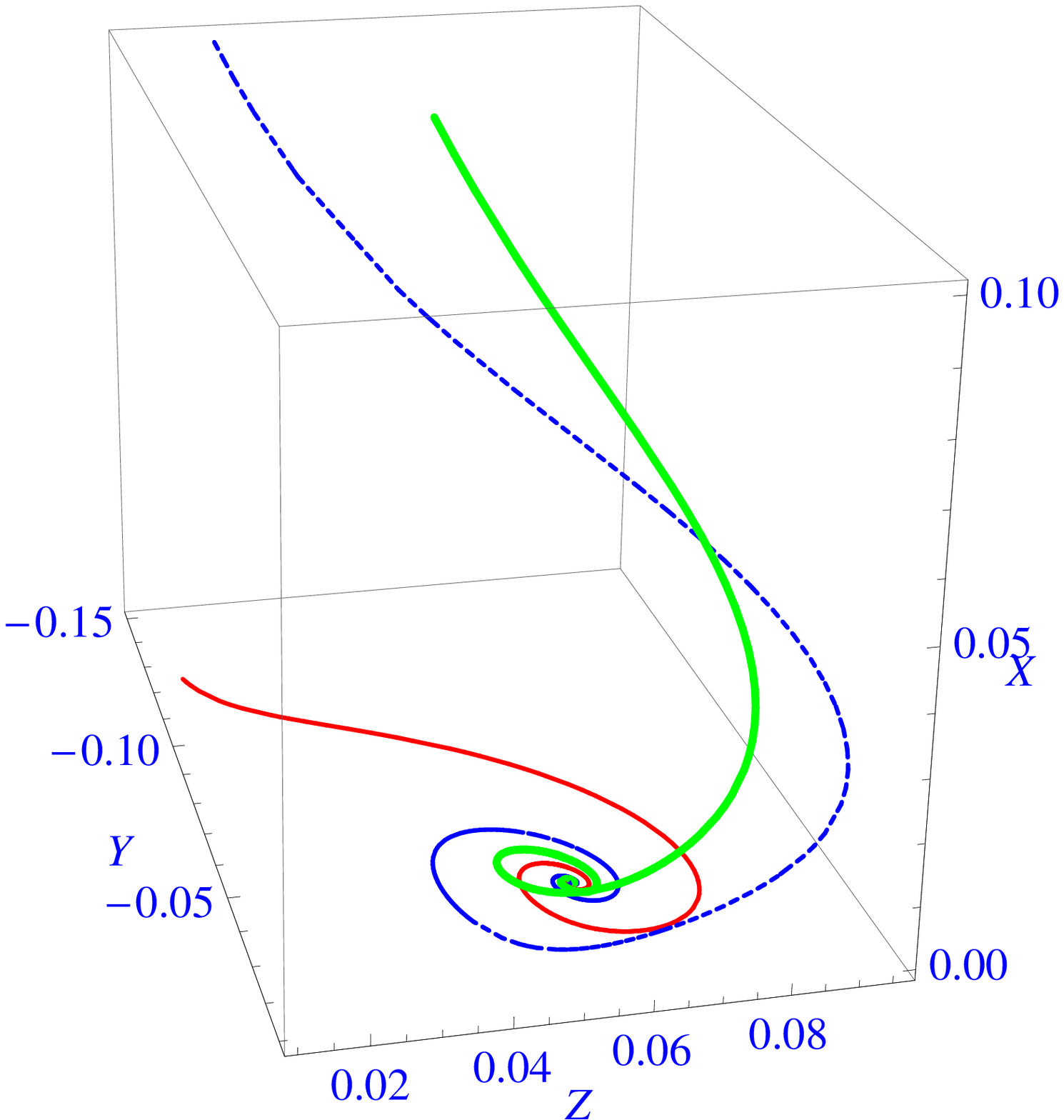}} \quad
{\includegraphics[height=40mm]{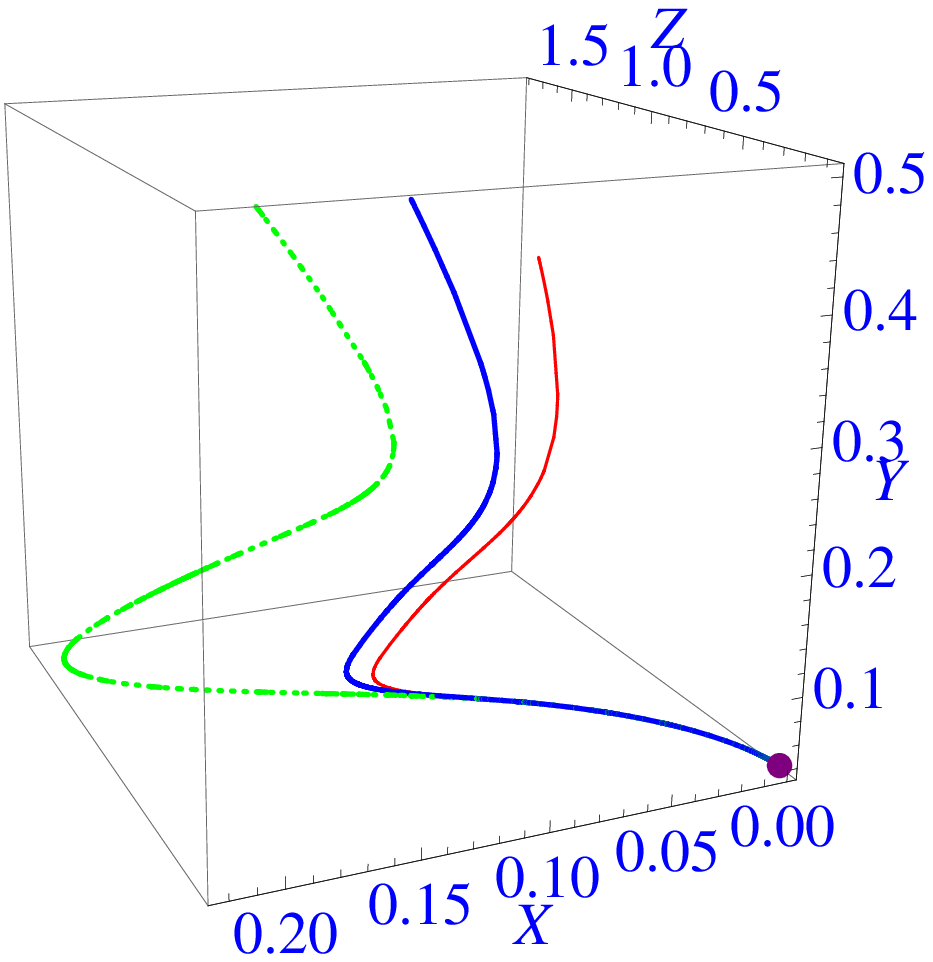}}
\end{center}
\caption{(From left to right) Attractor behavior of the anisotropic fixed points of the DBI, SDBI, and Galileon models, respectively. Here, the field parameters are chosen as $\lambda =0.1$, $\rho=50$, and $\gamma=1.5$ for the DBI model;  $\lambda =0.1$, $\rho=50$, and $\gamma=1.0001$ for the SDBI model; $\lambda =0.1$, $\rho=50$, and $k_0 =-3\rho^2/2$ for the Galileon model. Note that these pictures are taken from papers \cite{WFK1}, \cite{WFK2}, and \cite{WFK4}, respectively. } 
\end{figure}
\section{The role of phantom field to the validity of cosmic no-hair conjecture} \label{sec3}
We have shown above that the cosmic no-hair conjecture seems to be generally violated in the framework of KSW model, even when the scalar field is of non-canonical forms such as the DBI, SDBI, and Galileon fields, due to the existence of the unusual coupling term between scalar and vector fields, $f^2(\phi)F^2$. Hence, the validity of the cosmic no-hair conjecture in unusual scenarios might need the existence of unusual fields. Indeed, by introducing a phantom (scalar) field, whose kinetic energy is negative definite \cite{phantom}, we have been able to show that the corresponding anisotropic power-law inflationary solutions are indeed unstable rather than stable \cite{WFK}. This means that the validity of the cosmic no-hair conjecture really needs the existence of the phantom field in the KSW model \cite{WFK,WFK3} and its non-canonical extensions \cite{WFK1,WFK2,WFK4}. Now, we would like to briefly demonstrate here how unstable modes appear once the phantom field is introduced. Let us begin with an action of two-scalar-field extension of KSW model \cite{WFK}
\begin{equation}
  S = \int {d^4 } x\sqrt { - g}  \left[  {\frac{{M_p^2 }}
{2}R {-} \frac{{ 1 }}
{2} {\partial _\mu  {\phi} } {\partial ^\mu  {\phi} }  {+} \frac{{1 }}
{2} {\partial _\mu  {\psi} }  {\partial ^\mu  {\psi} }  - V_1({\phi})-V_2({\psi}) - \frac{1}
{4}f_1^2({\phi})  f_2^2({\psi}) {F_{\mu \nu } F^{\mu \nu }} } \right] . 
\end{equation}
As a result, the following field equations are given by 
\begin{eqnarray}
\ddot \phi & =&  - 3\dot \alpha \dot \phi  - \frac{{\partial V_1}}
{{\partial \phi }} + f_1^{ - 3}f_2^{-2} \frac{{\partial f_1}}
{{\partial \phi }}\exp\left[{ - 4\alpha  - 4\sigma } \right]p_A^2 ,\\
\ddot \psi  &= & - 3\dot \alpha \dot \psi  +\frac{{\partial V_2}}
{{\partial \psi }} - f_1^{ - 2}f_2^{-3}\frac{{\partial f_2}}
{{\partial \psi }}\exp\left[{ - 4\alpha  - 4\sigma } \right]p_A^2 , \\
 \dot \alpha ^2  &= &\dot \sigma ^2  + \frac{{1 }}
{3M_p^2}\left[ {\frac{1}
{2}\dot \phi ^2 -\frac{1}
{2}\dot \psi ^2  + V_1 +V_2 + \frac{f_1^{ - 2}f_2^{-2}}
{2} \exp\left[{ - 4\alpha  - 4\sigma } \right]p_A^2 } \right] , 
\\
\ddot \alpha  & =  &- 3\dot \alpha ^2  + \frac{{1 }}
{M_p^2} \left(V_1+V_2  \right) + \frac{{f_1^{ - 2}f_2^{-2}}}
{6M_p^2} \exp\left[{ - 4\alpha  - 4\sigma } \right]p_A^2 , 
\\
 \ddot \sigma & =  &- 3\dot \alpha \dot \sigma  + \frac{{f_1^{ - 2}f_2^{-2} }}
{3M_p^2} \exp\left[{ - 4\alpha  - 4\sigma } \right]p_A^2 .
\end{eqnarray}
And by choosing exponential forms of potentials and gauge kinetic functions such as \cite{WFK}
\begin{eqnarray}
V_1(\phi) &=& V_{01} \exp\left[{\frac{\lambda_1\phi }
{{M_p }} }\right] ;~ V_2(\psi)= V_{02} \exp\left[{\frac{\lambda_2\psi }
{{M_p }} }\right] ,\nonumber\\
f_1( \phi)f_2( \psi )& =& f_0 \exp \left[{\frac{\rho_1 \phi}{{M_p }} +\frac{\rho_2\psi }{{M_p }} }\right],
\end{eqnarray}
along with the following ansatz,
\begin{equation}
\alpha = {\zeta} \log \left( t \right);~ \sigma  = {\eta} \log \left( t \right); ~\frac{{\phi} }
{{M_p }}  ={ \xi_1} \log \left( t \right) + \phi _0;~ \frac{{\psi} }
{{M_p }}  ={\xi_2} \log \left( t \right) + \psi _0, 
\end{equation}
we have been able to figure out a set of anisotropic power-law solutions as \cite{WFK}
\begin{eqnarray}
{\zeta}  & =& \frac{{4\left( {\lambda _1 \rho _2  + \lambda _2 \rho _1 } \right)\left( { 2\lambda _1 \lambda _2+3\lambda _1 \rho _2  + 3\lambda _2 \rho _1 } \right)  + \lambda _1^2 \lambda _2^2 + 8\left( {\lambda _2^2  - \lambda _1^2 } \right)}}
{{6\lambda _1 \lambda _2 \left( {\lambda _1 \lambda _2  + 2\lambda _1 \rho _2  + 2\lambda _2 \rho _1 } \right)}}, \nonumber\\
{\eta} & =& \frac{{\lambda _1 \lambda _2 \left( { \lambda _1 \lambda _2+2\lambda _1 \rho _2  + 2\lambda _2 \rho _1   } \right) - 4\left( {\lambda _2^2  - \lambda _1^2 } \right)}}
{{3\lambda _1 \lambda _2 \left( {\lambda _1 \lambda _2  + 2\lambda _1 \rho _2  + 2\lambda _2 \rho _1 } \right)}}. \nonumber
\end{eqnarray}
And the corresponding inflationary solutions will approximately be
\begin{equation}
{\zeta} \simeq \frac{\rho_1}{\lambda_1}+\frac{\rho_2}{\lambda_2} \gg 1;~{\eta} \simeq\frac{1}{3},
\end{equation}
provided that $\rho_i \gg \lambda_i \sim {\cal O}(1)$. Next, we use {the power-law perturbations} \cite{WFK}, 
\begin{equation}
\delta \alpha, ~\delta \sigma,~\delta \phi, ~\delta \psi \propto t^n,
\end{equation}
which are compatible with the power-law solutions to study the stability of the obtained solutions. It is noted that $n>0$ and $n<0$ will correspond to unstable and stable perturbation mode(s), respectively. As a result, the following of equation of $n$ turns out to be  \cite{WFK}
\begin{equation}
{ f(n)} \equiv  {{n^7}+ b_7 n^6  + b_6 n^5  + b_5 n^4  + b_4 n^3  + b_3 n^2  + b_2 n  + { b_1}} = 0, 
\end{equation}
with
\begin{eqnarray}
 {b_1} &= &  - \frac{{2vl}}
{{\lambda _1 }}\left\{ { \left[ {\lambda _1^2 \lambda _2^2 \left( {5\zeta  - \eta  - 1} \right) + 2\lambda _1 \lambda _2 \left( {\lambda _1 \rho _2  + \lambda _2 \rho _1 } \right)\left( {3\zeta  - 3\eta  - 1} \right)} \right.} \right. \nonumber \\
 && \left. { + 4\left( {\lambda _1^2  - \lambda _2^2 } \right)} \right]\lambda _1u\left. { + 8\lambda _2 \rho _1 \rho _2 \left( {3\lambda _1 \rho _1 \zeta  - 3\lambda _1 \rho _1 \eta  - \lambda _1 \rho _1  - 2} \right)l} \right\},
\end{eqnarray}
where 
\begin{equation}
u=\frac{V_{01}}{M_p^2}\exp[\lambda_1 \phi_0];~v=\frac{V_{02}}{M_p^2}\exp[\lambda_2 \psi_0];~l=\frac{p_A^2}{M_p^2 f_0^2}\exp [-2(\rho_1 \phi_0 +\rho_2 \psi_0)].
\end{equation}
It is straightforward to see that $b_1 <0$ during the inflationary phase with $\zeta \gg \eta$, $u>0$, $v>0$, and $l>0$. Furthermore, we have observed in \cite{WFK} that $f(n=0)=b_1<0$, while $f(n)\sim n^7 >0$ as $n \gg 1$. This indicates that the curve $f(n)$ will cross the positive horizontal axis at least one time, i.e.,  the equation $f(n)=0$ will admit at least one positive root $n>0$ \cite{WFK}. Hence, the corresponding anisotropic power-law solution is really unstable during the inflationary phase as expected. Note that the same conclusion can be found in the other non-canonical extensions of KSW model, i.e., the DBI, SDBI, and Galileon models \cite{WFK1,WFK2,WFK4}, as well as in an extension of two-scalar-field model, in which a mixed kinetic term of canonical and phantom fields, $\partial_\mu \phi \partial^\mu \psi$, is involved \cite{WFK3}.   
\section{Conclusions}\label{sec4}
We have presented basic details of our recent studies on the fate of the cosmic no-hair conjecture in the context of KSW model. In particular, we have shown that the KSW model always admits the Bianchi type I metric as its power-law stable and attractor solutions for both canonical and non-canonical scalar fields due to the existence of the unusual coupling term between the scalar and vector fields. This indicates that the Hawking cosmic no-hair conjecture seems to be generally violated in the context of KSW model. However, we have also shown that the conjecture might be valid if the phantom field with its negative kinetic term is involved into the KSW model. As a result, the existence of phantom field will introduce at least one unstable mode to field perturbations, and therefore any initial spatial anisotropy of Bianchi type I metric will decay as the cosmic no-hair conjecture states. Note that this conclusion has been shown to be valid for all non-canonical extensions of KSW model mentioned above \cite{WFK1,WFK2,WFK3,WFK4}. Hence, we come to a conclusion that the phantom field might play an important role in order to ensure the validity of the cosmic no-hair conjecture. 
\ack The author is deeply grateful to Professor W. F. Kao of  Institute of Physics in  National Chiao Tung University for his useful advice on the KSW anisotropic inflation model and the cosmic no-hair conjecture. The author would like to thank Dr. Ing-Chen Lin as well as Dr. Sonnet Hung Q. Nguyen very much for their collaboration. This research is supported in part by  the Vietnam National Foundation for Science and Technology Development (NAFOSTED) under Grant No. 103.01-2017.12.

\end{document}